\documentclass[11pt,a4paper]{article}

\pdfoutput=1

\usepackage{jheppub}

\date{\today}

\usepackage{amsmath}
\usepackage{amsfonts}
\usepackage{amssymb}
\usepackage{ifthen}
\newcommand{\sech}{\mathop{\mathrm{sech}}}
\newcommand{\re}[1]{(\ref{#1})}
\newcommand{\arc}{\mathop{\mathrm{arc}}}

\begin{document}

\title{Oscillons in the presence of external potential}

 \author[a]{Tomasz Roma\'nczukiewicz}

\author[b,c,d]{Yakov Shnir}
\affiliation[a]{Institute of Physics, Jagiellonian University,\\
\L ojasiewicza 11, 30-348 Krak\'ow, Poland}
\affiliation[b]{BLTP, JINR,\\Dubna 141980, Moscow Region, Russia}
\affiliation[c]{Department of Theoretical Physics and Astrophysics,BSU,\\ Minsk 220004, Belarus}
\affiliation[d]{Department of Theoretical Physics, Tomsk State Pedagogical University,\\ Russia}

\emailAdd{trom@th.if.uj.edu.pl}
\emailAdd{shnir@maths.tcd.ie}

\abstract{We discuss similarity between oscillons and oscillational mode in perturbed $\phi^4$. For small depths of the perturbing potential it is
difficult to distinguish between oscillons and the mode in moderately long time evolution, moreover one can transform one into the other by adiabatically switching on and off the
potential. Basins of attraction are presented in the parameter space describing the potential and initial conditions.}
\keywords{Field Theories in Lower Dimensions, Nonperturbative Effects, Solitons Monopoles and Instantons}
\maketitle

\section{Introduction}
In the field theory the long time evolution of a dynamical system is usually governed by the
spectral structure of small perturbations around the static solution,
such as for example a soliton. There the eigenmodes of the corresponding linearized problem
play very important role. Usually there is a continuous
spectrum of scattering modes and a discrete sequence of oscillational (normal) modes. The lowest mode of the continuum
is often referred to as the mass threshold.

The resonance states, or quasinormal modes also play quite important role in the long time dynamics of the system.
They are defined as the solution of the linearized system which corresponds to the purely outgoing wave, they
can also be defined as the complex poles
of the Green function. Thus, their eigenfrequencies are complex valued,
the imaginary part of them is responsible for exponential decay of the corresponding mode.
Usually the mode with the lowest damping is the one which is seen in the late time dynamics, it is the most long living.

Recall that in the linear system the normal modes, which have real frequencies, live infinitely long.
However, in nonlinear theories these modes are coupled to the states of the continuum spectrum.
In the case of the quadratic non-linearity such a coupling yields an outgoing wave which has twice the frequency
of the oscillational normal mode. Such a wave carries away the energy and causes the oscillational mode to decay.
As shown in \cite{Manton:1996ex,Slusarczyk:1999iy}, this results in the
decay of the discrete mode amplitude according to the power law, $A\sim t^{-1/2}$.

However, if the oscillational mode has a low frequency, even the corresponding second harmonic
may still be below the mass threshold and therefore it cannot propagate. Then
only the higher harmonics can carry away the energy. This possibility was firstly described in
\cite{Dorey:2015sha}.

Note that the nonlinearities also can lower the frequency of the oscillational mode when it is
highly excited.
However, when its amplitude is decreasing, the corresponding frequency tends to the value one can find
from the linearized problem.

Usually, the scattering modes of the continuous spectrum quickly radiate the energy away
(except the quasinormal resonance  modes) and do not play any role in the long time
evolution of the system. However, there is yet another exception.
Recall that the speed of propagation of waves in massive field theories depends on the frequency. The slowest modes are
the ones which are near the mass threshold when the propagation frequency tends to zero. As a result
the generic initial conditions may develop slowly
decaying field with the frequency of the threshold. As was shown for example in \cite{Bizon:2011zz}
this oscillations decay as $t^{-1/2}$ or, when a resonance is on the threshold, as $t^{-2/3}$.

Peculiar feature of many non-linear models is that they also support time dependent
non-perturbative soliton-like solutions which are not
captured by the linear analysis. The well know example is the breather configuration in
the 1+1 dimensional sine-Gordon model. It is an exact periodic solution to the nonlinear
equation which does not lose its energy  into radiation, so it has an infinite lifetime.
However, sine-Gordon model is exceptional since it is integrable and the states of the continuum are completely
separated from the solutions of the field equations.
On the other hand, in non-integrable models, like $\phi^4$ theory, such a  breather-like configuration
with finite energy cannot exist infinitely long
\footnote{The model with V-shaped potential \cite{Arodz:2007jh} may support non-radiating
time-periodic compacton solution, which is an exception.} because they  decay radiating small amount of
energy away \cite{Fodor:2008du}.

However, in the $\phi^4$ model there are quasi-non-dissipative and almost periodic time-dependent configuration, the
oscillons \cite{Bogolyubsky:1976nx,Copeland:1995fq,Gleiser:1993pt}. They
appear as quasi-breathers, which are localised, extremely long-lived regular
finite energy configurations  performing non-harmonic oscillations about the one of the vacua.
Typically,  the radiation energy of the oscillon is extremely small, numerical simulations reveal that in 1+1 dimensions
the oscillon survives after a few millions of oscillations \cite{Fodor:2006zs,Salmi:2012ta} with the decay rate
$dE/dt\sim-\exp(-B/E)$ beyond all orders \cite{Fodor:2008du}.

The interplay between the states of the perturbative spectrum of the $\phi^4$ theory and solitons, kinks and oscillons,
attracted a lot of attention. The most familiar observation is the resonance effect observed in the bouncing collisions of
the kink--anti-kink ($\bar KK$)  pair  and excitation of the internal
vibrational mode of the kink \cite{Campbell:1983xu,Anninos:1991un}.

It was pointed out that the excitations of the internal mode of the kink may produce
kink--anti-kink pairs \cite{Manton:1996ex}. In \cite{Romanczukiewicz:2005rm} it was shown that such mode can be excited by radiation via
parametric resonance leading to the defect production. The kink--anti-kink pair can also be
produced in the absence of a single soliton via the
resonance excitations of the oscillon by the incoming waves \cite{Romanczukiewicz:2010eg}
showing another resemblance between oscillons and
oscillational modes.
%

In the present paper we will show that there is also a correspondence between the internal mode of the external potential
and the oscillon, which under special conditions, can smoothly be transferred into each other.


\section{The model}
\subsection{Definition and static solutions}
We consider the standard one-dimensional  $\phi^4$ model with an additional external potential $V(x)$ defined by Lagrangian
density
\begin{equation}
 \mathcal{L}=\frac{1}{2}\phi_t^2-\frac{1}{2}\phi_x^2-\frac{1}{2}\left(\phi^2-1\right)^2-\frac{1}{2}V(x)(\phi-1)^2
 \label{lag}
\end{equation}
As the perturbation potential we have chosen the P\"oschl-Teller potential
\begin{equation}
 V(x)=-\frac{V_0}{\cosh^2 bx}
\label{Poeschl-Teller}
\end{equation}
since it is vanishing asymptotically and all solutions of the corresponding eigenvalue problem are known analytically.
Because oscillons are symmetric we limit our considerations only to even symmetry $\phi(-x)=\phi(x)$.

Physically, such a trapping potential corresponds to a localized inhomogeneity
(impurity), similar impurity potential was introduced in the sine-Gordon models to
investigate the effects of inhomogeneities in the kink-antikink scattering process \cite{Fei:1992dk,Goodman}.
In the limiting case of a localized impurity at the origin, the defect corresponds to a delta-function, such
a potential was used in the $\phi^4$ model to study kink-impurity interactions \cite{Kivshar:1991zz},
further investigation reveal very
interesting pattern of resonance scattering of the kinks on the impurities \cite{Goodman}.

The static solutions obey the time-independent equation
\begin{equation}
 \phi_{xx}=2\phi\left(\phi^2-1\right)+V(x)(\phi-1).
\end{equation}
For even symmetry the solutions has no topological charge $\phi(-\infty)=\phi(\infty)=\phi_v$, where $\phi_v=\pm1$ is one of the usual $\phi^4$
vacuum. The second condition is $\phi_x(0)=0$.
These conditions, however, do not define unique solutions.
\subsubsection{Sector $\phi=1$}
In the topological sector with  $\phi_v=1$ there are two solutions. First one is the trivial vacuum
$\phi\equiv1$, clearly this is zero energy state.

There is yet another solution in this sector,
which for small values of $V_0$ resembles an antikink--kink pair ($\bar KK$)
\begin{equation}\label{eq:KK1}
 \phi(x)\approx \tanh(x-X)-\tanh(x+X)+1.
\end{equation}
Energy of such configuration for $b=1$, relatively small values of $V_0$ and large values of $X$,
can be found approximately
\begin{equation}\label{eq:EnergyStat1}
 E\approx2M-16e^{-4X}-2\pi V_0\left(1-2e^{-X}\right),
\end{equation}
where $M=4/3$ is a mass of a single kink. This is the usual Yukawa interaction between the well separated kink antikink pair,
modulated by the perturbation potential.
Since the Yukawa interaction is attractive, the solution of that type
cannot exist in the absence of the external potential, nor for $V_0<0$ when
the potential also becomes attractive.

\begin{figure}[!h]
\centering
\includegraphics[width=.75\linewidth]{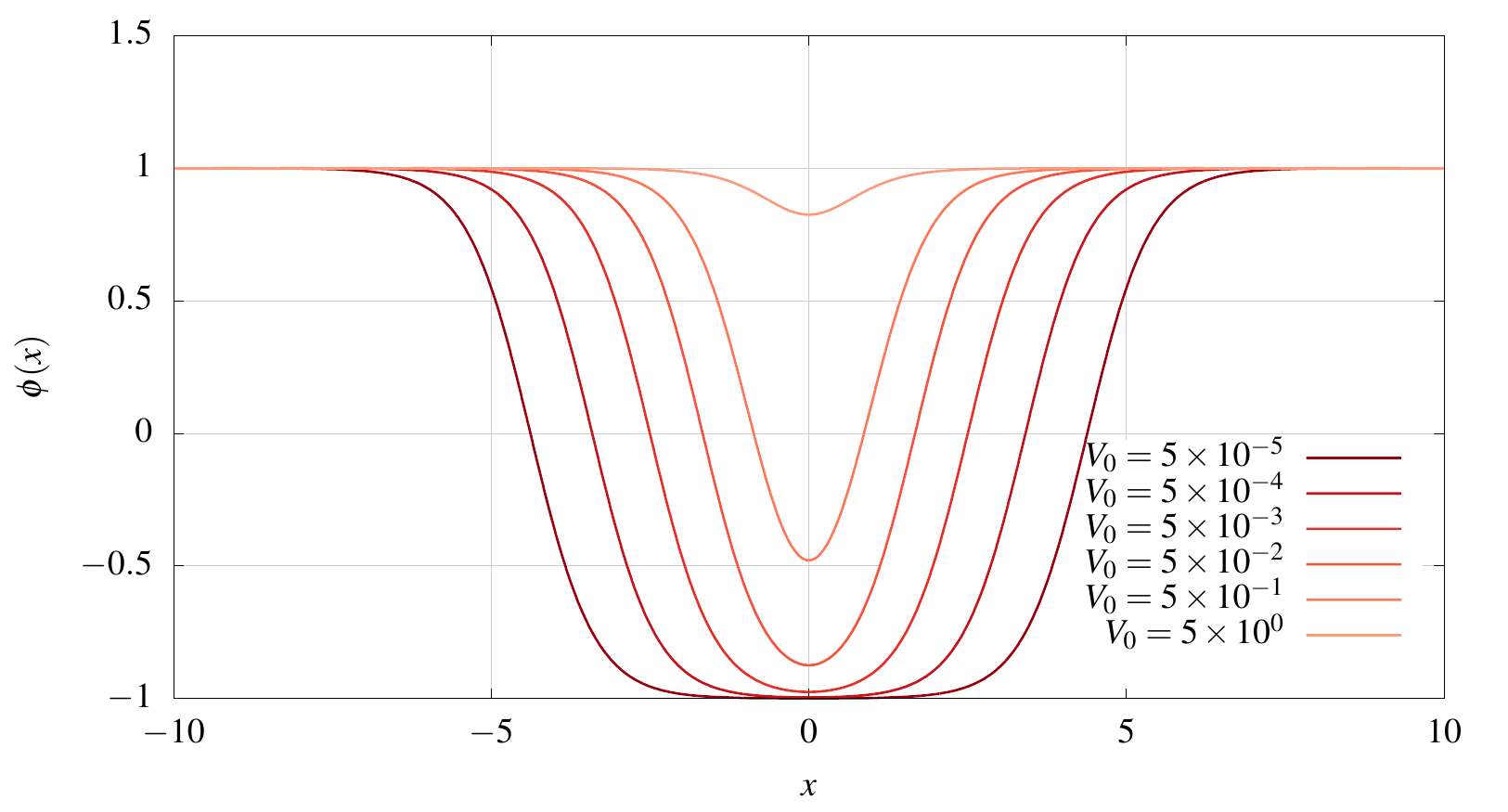}
   \caption{\small Static (unstable) $\bar KK$ solutions  in the topologically trivial sector.}\label{fig:staticProfiles2}
\end{figure}

However, for $V_0>0$ the energy for $\bar KK$ has a maximum for
\begin{equation}
 X=-\frac{1}{3}\log\left(\frac{\pi V_0}{16}\right).
\end{equation}
Thus, the pair $\bar KK$ can be held together by the external potential,
however, small perturbation can destabilize the configuration as it is the
maximum of energy and the pair can either separate or collapse and annihilate.
As $V_0$ increases the approximation (\ref{eq:KK1}) is no longer valid and the shape of the solution becomes a lump centered at $x=0$ vanishing for
$V_0=6$.
Energy of this solution decreases from the energy of two static kinks as $V_0=0$: $E(0)=2M=8/3$ to 0 as $V_0=6$.
In Fig.~\ref{fig:staticProfiles2} we display some example of corresponding profiles
for some set of values of $V_0$. For very small value of $V_0=5\cdot10^{-5}$ we can evaluate the separation between the kinks from
the equation (\ref{eq:EnergyStat1}), it is equal to $X=3.84$ which seems to be just little less then
the value obtained from the solution of the corresponding static equation.

\subsubsection{Sector $\phi=-1$}

Note that unlike the original $\phi^4$ model, the model \re{lag} has only one trivial solution $\phi=1$.
In the second topological sector the static configuration,  which asymptotically tends to $\phi=-1$ as $x\pm
\infty$, is a nontrivial solution localized by the potential \re{Poeschl-Teller}.
It is similar to a non-topological soliton (lump), which appears in the
two-component system of coupled fields \cite{Rajaraman:1978kd,Halavanau:2012dv}.
However, in the model \re{lag} the lump centered at $x=0$
is trapped by the potential $V(x)$, it cannot
propagate.

\begin{figure}[!h]
 \begin{center}
\includegraphics[width=.75\linewidth]{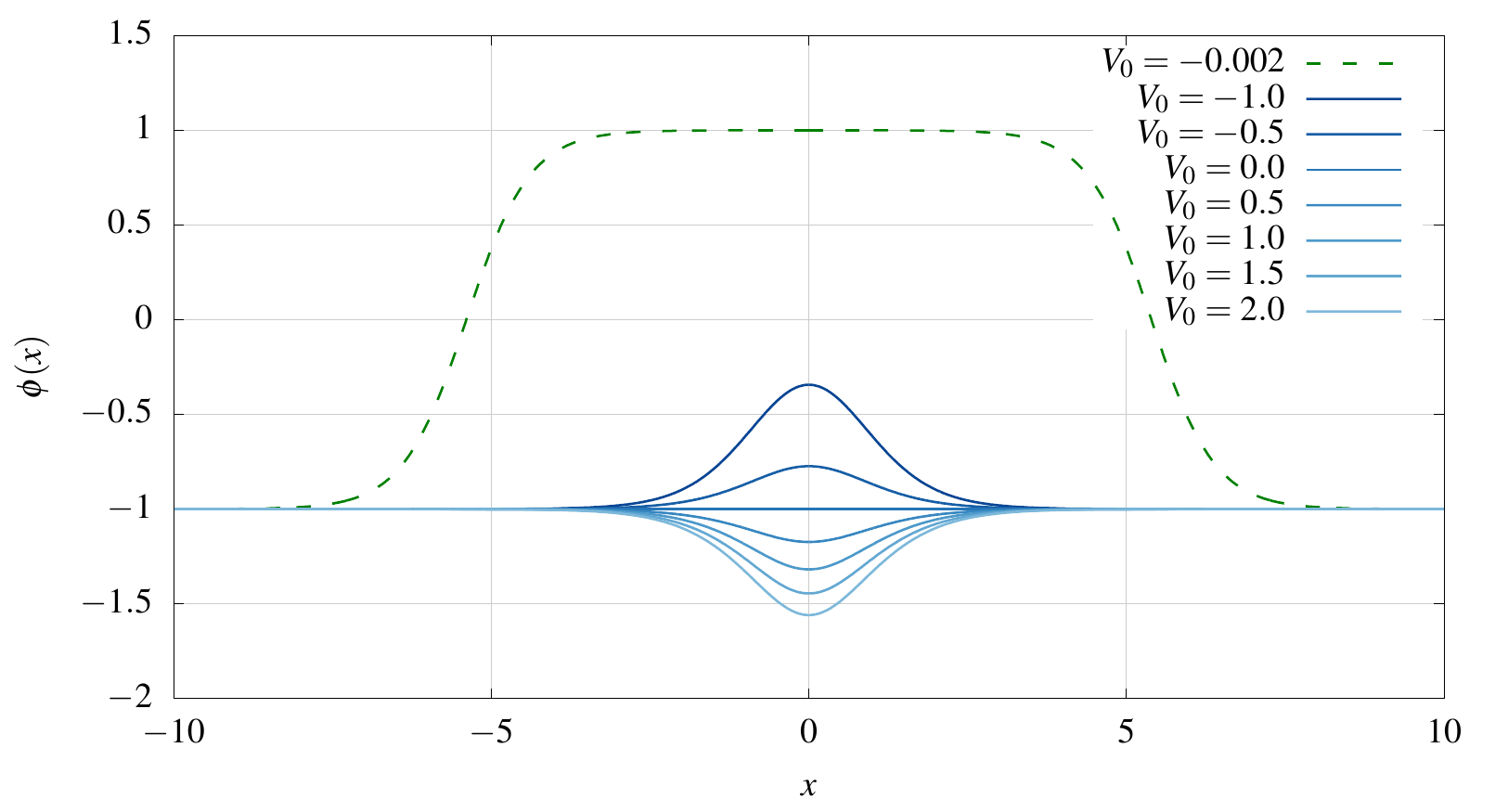}
\caption{\small Static lump solutions of the model \re{lag} with the potential \re{Poeschl-Teller}
at $b=1$ for a few different values of $V_0$.}\label{fig:staticProfiles}
\end{center}
\end{figure}

Although we have not found this solution analytically, one can easily find it  numerically, in
Fig.~\ref{fig:staticProfiles} we plotted the corresponding profiles of the vacuum configurations
at $b=1$ and a few values of the parameter $V_0$. As $V_0$ is positive, $\phi(0) < -1$, for
negative values of the parameter $V_0$ the value of the field at the center of the trapping potential
is  $\phi(0) > -1$. The energy of the lump is set to be zero as $V_0=0$, it is positive for negative
values of $V_0$ and it is negative as $V_0 >0$ decreasing monotonically as
$V_0$ increases.

Again, as in the first sector, we have found another solution, which represents a static
kink-antikink pair, captured by the potential
\begin{equation}
 \phi(x)\approx \tanh(x+X)-\tanh(x-X)-1.
\end{equation}
The energy of this configuration, for large value of $X$ and small $V_0$ can be found as
\begin{equation}
 E\approx2M-16e^{-4X}-4\pi V_0e^{-X}.
\end{equation}
The above energy has maximum for
\begin{equation}
 X=-\frac{1}{3}\log\left(-\frac{\pi V_0}{16}\right).
\end{equation}
is just above the total mass of two static $\phi^4$ kinks, $E\approx 2 M = 8/3$.
Clearly  this solution is unstable, it can decay into the vacuum $\phi=1$
with two kinks escaping to infinity or into the lump solution which is in the same topological sector but has less energy.

The stable lump solution and a single example of the unstable $K\bar K$ solution are shown in Figure \ref{fig:staticProfiles}.

\begin{figure}[!h]
\includegraphics[width=.75\linewidth]{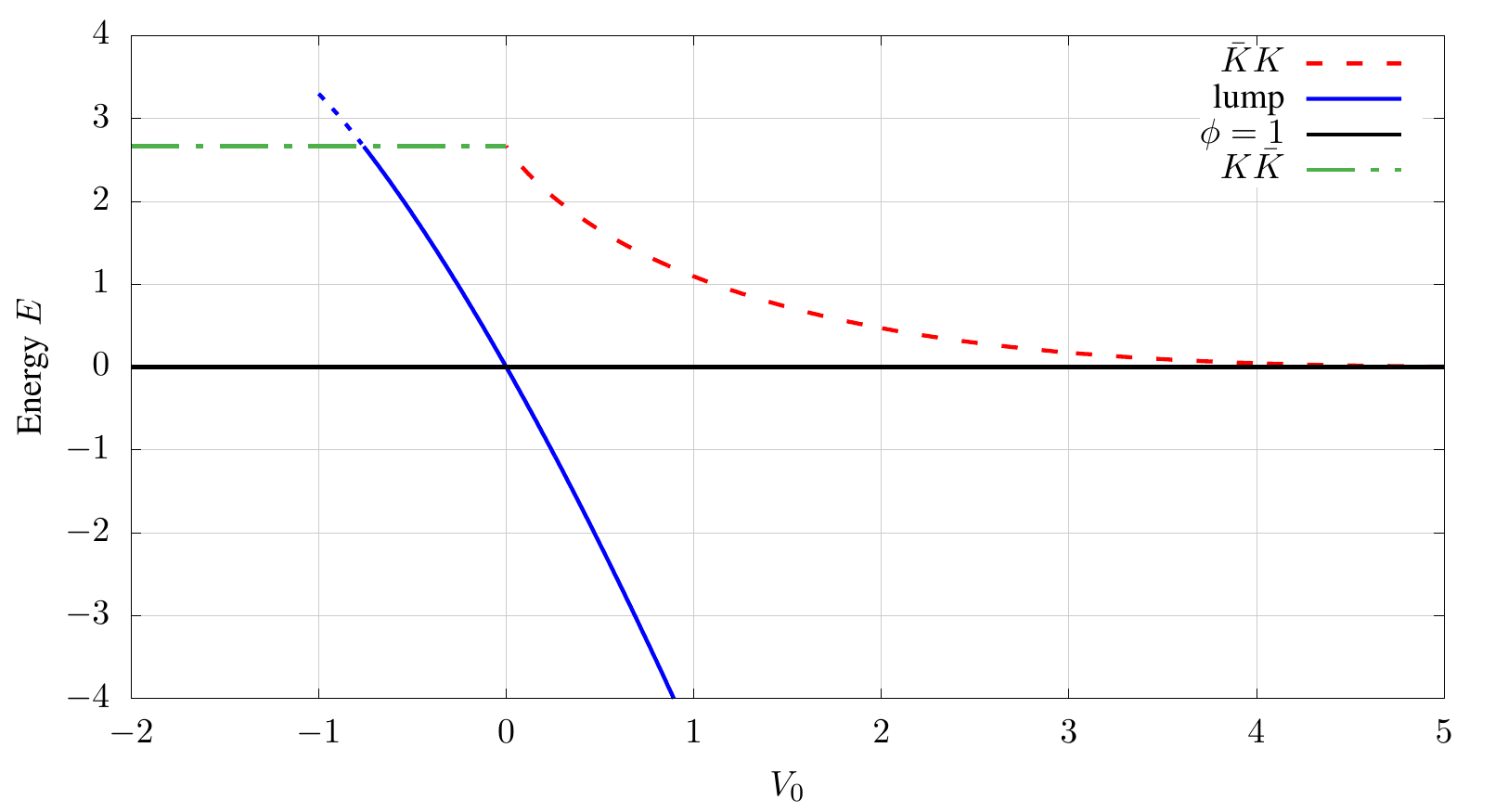}\centering
   \caption{\small Energies of the static solutions of the model
\re{lag} vs $V_0$. Dashed curves correspond to the unstable solutions, see
Figs.~\ref{fig:staticProfiles2} and \ref{fig:staticProfiles} }\label{fig:staticEnergies}
\end{figure}

The energies of the four static solutions are gathered in the Figure \ref{fig:staticEnergies}.
Note that the energy of a lump increases as $V_0$ decreases. For $V_0<-0.761$ the energy of the lump is larger then the energy of the unstable $K\bar
K$. This means that the system has no stable solution in the sector $\phi(|x|\to\infty)=-1$,  it always decays
to the $\phi=1$ solution emitting a kink-antikink pair.

For $V_0>0$ the lump solution has lower energy than the stable solution in the first topological sector,
$\phi=1$. However, for $V_0<6$  the system
cannot jump spontaneously from $\phi=1$ to the lump solution emitting the kink-antikink pair because of the energy threshold.
The minimal energy which
allows such a transition is defined by the mass of the static $\bar KK$ configuration. Note the energy difference
between the lump solution  and the pair $\bar KK$, quickly
increases as $V_0$ grows. Hence the velocity of the emitted pair of  kinks also increases.

In the rest of the paper we will consider the
excitations of the solution in the sector $\phi=1$.
So the process of $\bar KK$ emission would be more and more possible as $V_0$ grows and the energy threshold
decreases to 0 for $V_0=6$.

\subsection{Linear stability of static solutions}

Generally, solutions of a nonlinear field theoretical model can be decomposed into the infinitely many modes
oscillating with different frequencies.
However, our study focuses on some special solutions which are almost periodic.
Such a period defines the basic frequency of the solution, which we would refer to as the frequency.
The power spectrum then usually consists of the basic frequency along with the higher harmonics, which,
due to nonlinearity, are often contaminated via excitations of other modes. During the time evolution the
basic frequency may change, typically the corresponding time interval is much larger than
the period of the solution. On the other hand, we can consider linearized equations of motion, in such a case the
frequency of the corresponding oscillational modes, would be referred to as the linearized frequency or eigenfrequency.

Let us consider the linearized fluctuations around the trivial vacuum $\phi=1+e^{i\omega t}\eta(x)+c.c.$
in the model \re{lag}. The corresponding eigenfunctions are solutions of the equation
\begin{equation}
-\eta_{xx}+\left(m^2+V(x)\right) \eta=\omega^2\eta.
\label{linear-eq}
\end{equation}
where $m=2$ is the mass of the mode. The potential (\ref{Poeschl-Teller}) yields the ground state
\begin{equation}\label{eq:linsol}
 \eta_0(x) = \frac{1}{\cosh^{\lambda}(bx)}, \, \, \,  \lambda(\lambda+1)=\frac{V_0}{b^2},\;\,\, \, \omega^2=m^2-b^2\lambda^2
\end{equation}
where $\lambda$ is a  real parameter. Recall that in the particular case when $\lambda$ is an integer,
the P\"oschl-Teller potential is reflectionless.

\begin{figure}[!h]
 \begin{center}
\includegraphics[width=.99\linewidth]{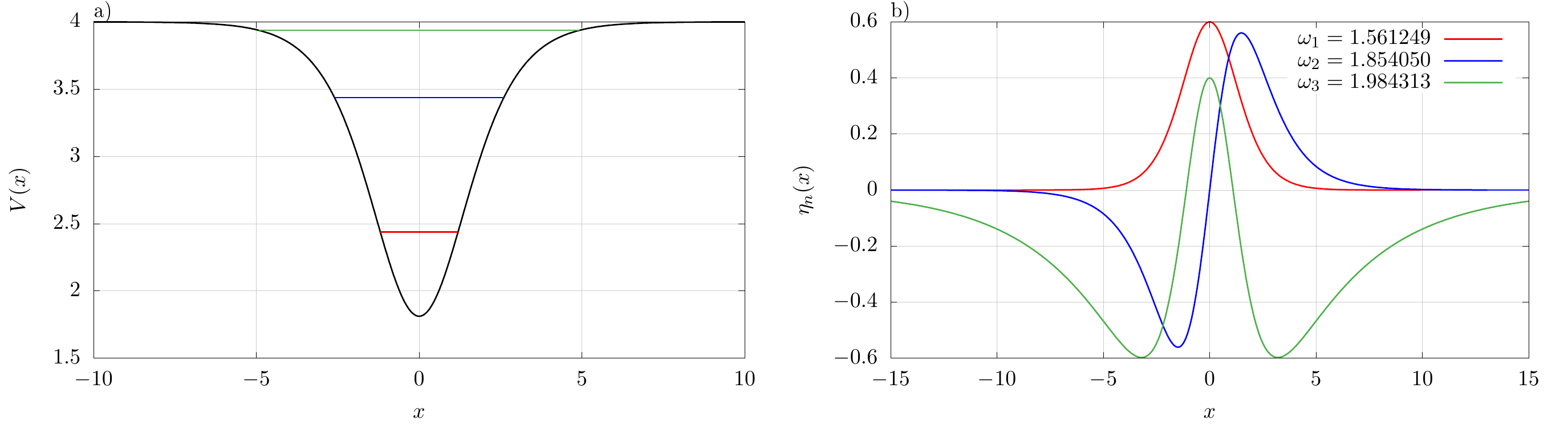}
\caption{\small Example potential (a) and the profiles of the bound states (b) for $b=0.5$, $\lambda=2.5$ and
$V_0=2.1875$.}\label{fig:Potentials}
\end{center}
\end{figure}
\begin{figure}[!h]
 \begin{center}
\includegraphics[width=.75\linewidth]{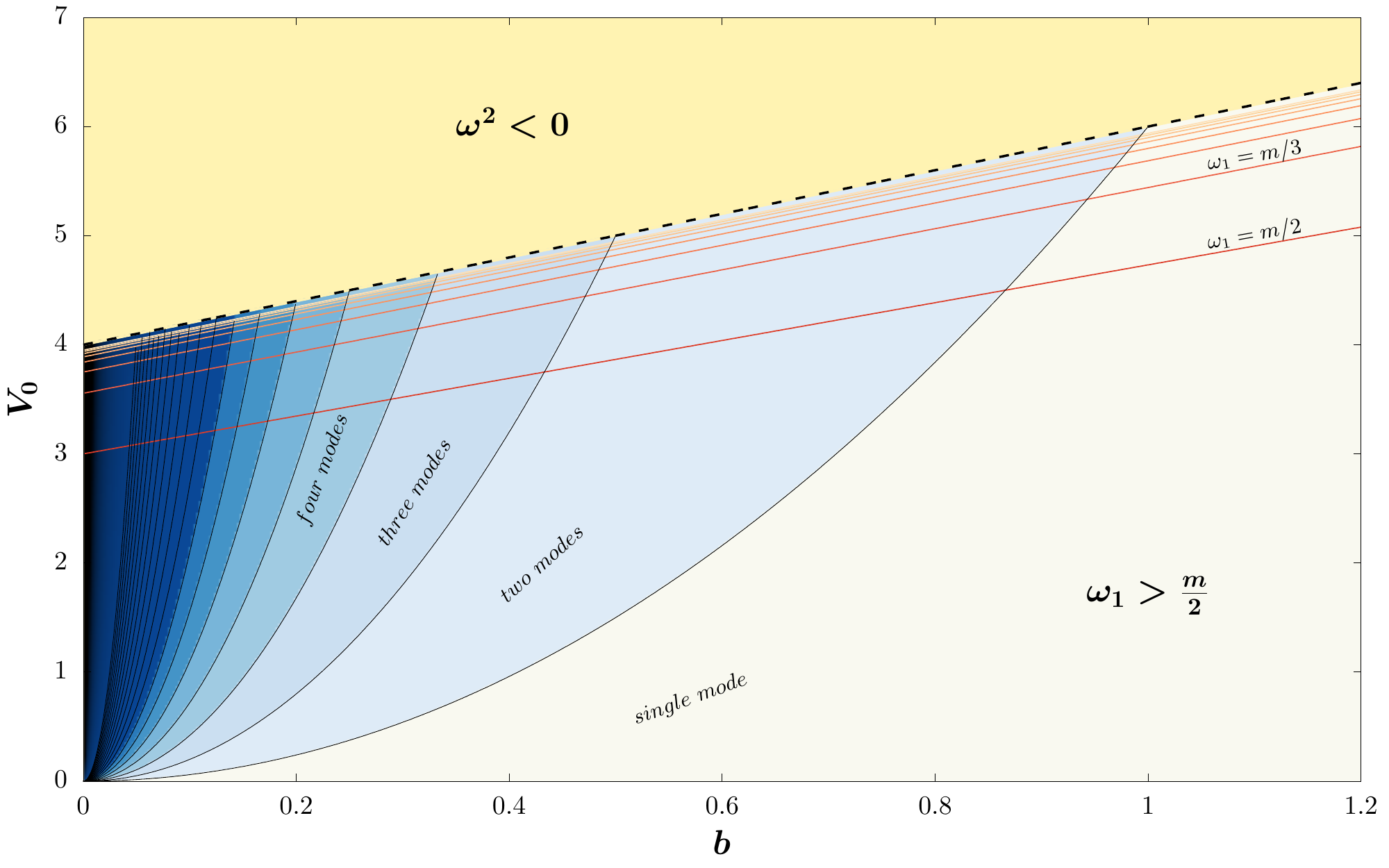}
\caption{\small Spectral structure of even solutions of P\"oschl-Teller potential.
Red and orange lines correspond to values where the frequency of the first mode is equal to $m/n$ for integer
$n\geq2$.}\label{fig:smallPerturbation}
\end{center}
\end{figure}

The equation for $\lambda$ has exactly one positive solution for $V_0>0$.
For $V_0>4+2b$ the eigenfrequency of the ground state becomes imaginary, i.e. $\omega^2<0$ (Figure \ref{fig:smallPerturbation}).
It means that the corresponding mode is unstable and the system could
change its ground state producing kink--anti-kink pair.
Further, one can expect that in the presence of the
perturbation potential, the lowest oscillational mode could become an attractor in the
time evolution of some even initial data, which, in the limit $V_0=0$ would evolve into the oscillon-like state.
For $b<1$ there can be more eigensolutions, frequencies of which are given by
\begin{equation}
 \omega_j^2=m^2-b^2(\lambda-j)^2,\qquad 0\leq j < \lambda.
\end{equation}
Similar analysis around the second solution, which asymptotically approaches the vacuum $\phi=-1$ as $x\to \pm
\infty$, shows that there are no oscillational modes for $b=1$ and $V_0<0$.
The corresponding effective potential, which appears in the linearized equation for excitations is
\begin{equation}
 V_{eff}(x)=6\phi_s^2(x)-2+V(x)\, ,
\end{equation}
it is repulsive for $V_0<0$. $\phi_s^2(x)$ is the static solution.
Below we mainly restrict our
consideration to the case of fluctuations about the trivial vacuum $\phi_v=1$.

\subsection{Radiative decay of the oscillational modes}
In nonlinear theories the oscillational modes are coupled to the higher frequency modes
from the continuous spectrum.
These higher modes can propagate radiating away the energy, initially stored in the oscillational modes, and causing
their decay.  In particular, it has been shown that the rate, at which
the oscillational mode of the $\phi^4$ kink decays, obeys the power law \cite{Manton:1996ex}
\begin{equation}
 A(t)\sim t^{-1/2}.
\end{equation}
In the Appendix we show that the similar law can be applied to the shallow oscillational modes of the
external potential in the model \re{lag}.
Note that there is a single oscillational mode of the $\phi^4$ kink configuration, it decayed via
radiation propagating through the second harmonic.
On the other hand, the model \re{lag} may also support higher oscillational modes with eigenfrequencies below $m/2$
for which the second harmonic cannot propagate because it is below the mass threshold.
In such a case the decay is caused by at least the third harmonic generated in the third order of the perturbation scheme.
This leads to the different character of the decay.
In the Appendix  we provide more details of the process of decay of the higher oscillational modes in the model
 \re{lag} and discuss the dynamical transitions between different scenarios.

Throughout the paper (except the Appendix) we will restrict our considerations to the case of $V_0<2$, this corresponds
to the oscillational modes with frequencies well above the threshold $m/2$, above which the second harmonics can propagate.

As the oscillational mode is excited to a nonlinear regime, its frequency becomes lowered and the profile becomes more narrow.
Evolving in time, such a mode radiates some amount of energy, so its amplitude decreases according to the Manton-Merabet
rule $A\sim t^{-1/2}$ \cite{Manton:1996ex}. The frequency of this mode increases with time, its profile becomes
broader as the mode evolves toward the linear regime.

\subsection{Oscillons}
The lifetime of the oscillons is very large, so they
 can be well approximated by the Fourier decomposition
\begin{equation}
 \phi(x,t) = 1+\sum_{n=0}^N\phi_n(x)\cos(n\omega t)\, .
\end{equation}
Note that such a periodic solution with the frequency $\omega<m$
cannot have a finite energy\footnote{The sine-Gordon breathers, which are
solutions of the integrable model, are exceptions.}. We can consider the
pseudobreathers, which are the corresponding oscillating solutions, which
minimize the oscillatory tails. Such tails represent a standing wave which can be considered as a superposition
of two waves, propagating in opposite
directions. One of these waves can be considered as the radiation, which carries away the energy, whereas
the other incoming wave stabilizes the pseudobreather.

However, the physical oscillons possess only outgoing radiation tails.
During the evolution the amplitude of the oscillon slowly decreases, and the rate, in which it
radiates, decreases even faster. Therefore oscillons
live exceptionally long and the decay rate is beyond all orders \cite{Fodor:2008du}
\begin{equation}
 \frac{dE}{dt}\sim\exp(-B/E)\,.
\end{equation}

In the unperturbed 1+1 dimensional $\phi^4$ the oscillons tend to the lowest mode of the continuous spectrum at $\omega=m$.
As it was shown in \cite{BOYD1995311,Fodor:2008du}, the $\phi^4$ model can be treated as a perturbation to the sine-Gordon model
\begin{equation}
 u_{tt}-u_{xx}+\sin u=0\,,
\end{equation} which exhibits exact, periodic solutions called the breathers
\begin{equation}
 u(x,t)= 4\arc\tan\left[\frac{\sqrt{1-\omega^2}}{\omega}\frac{\cos(\omega t)}{\cosh\left(\sqrt{1-\omega^2}\,x\right)}\right]\,.
\end{equation}
To match the expansion of the sG equation with the $\phi^4$ equation around $\phi=1$ one has to rescale the variables
$(x,t,\omega)_{sG}\to(2x, 2t, \omega/2)_{\phi^4}$.
It is conjectured that the first approximation to the oscillon can be written as a rescaled breather
\begin{equation}
 \phi(x,t)\approx1+4\arc\tan\left[\frac{\sqrt{1-\omega^2/4}}{\omega/2}\frac{\cos(\omega t)}{\cosh\left(2\sqrt{1-\omega^2/4}\,x\right)}\right]
\end{equation}
which should be valid for low amplitude oscillons.
Defining the new variable
\begin{equation}
 \epsilon=2\sqrt{1-\omega^2/4}
\end{equation}
the lowest approximation of their profiles in $\phi^4$ can be nicely described as
\begin{equation}\label{oscapprox}
 \phi_1(x)=2\epsilon\sech(\epsilon x)\, ,
\end{equation}
which oscillates with the frequency $\omega=2\sqrt{1-\epsilon^2/4}$.
Other profiles and the correction to the profile of $\phi_1$ are of order $\mathcal{O}(\epsilon^2)$.
It is worth mentioning that (\ref{oscapprox}) is not a solution of any linearized equation and higher order corrections (even though much smaller
than $\phi_1$) are essential to stabilize the solution.

\section{Numerical Results}
\subsection{Numerical methods}
To solve the equation of motion following from the Lagrangian (\ref{lag})
we applied the method of lines using five point stencil to approximate the
spatial derivative and fourth order symplectic method to integrate in time.
Typically the space grid was $\delta x=0.05$ and time step $\delta t=0.4\delta x$, but we change the values to ensure stability and convergence of the
solutions. At $x=0$ we imposed symmetric boundary conditions $\phi(-x)=\phi(x)$. The second boundary was chosen in such distance (more than half of
the evolution time) so that
the waves had no time to reflect and come back to interfere with the field at $x=0$.
\subsection{Oscillon and mode evolution}
\begin{figure}[!h]
 \begin{center}
\includegraphics[width=.75\linewidth]{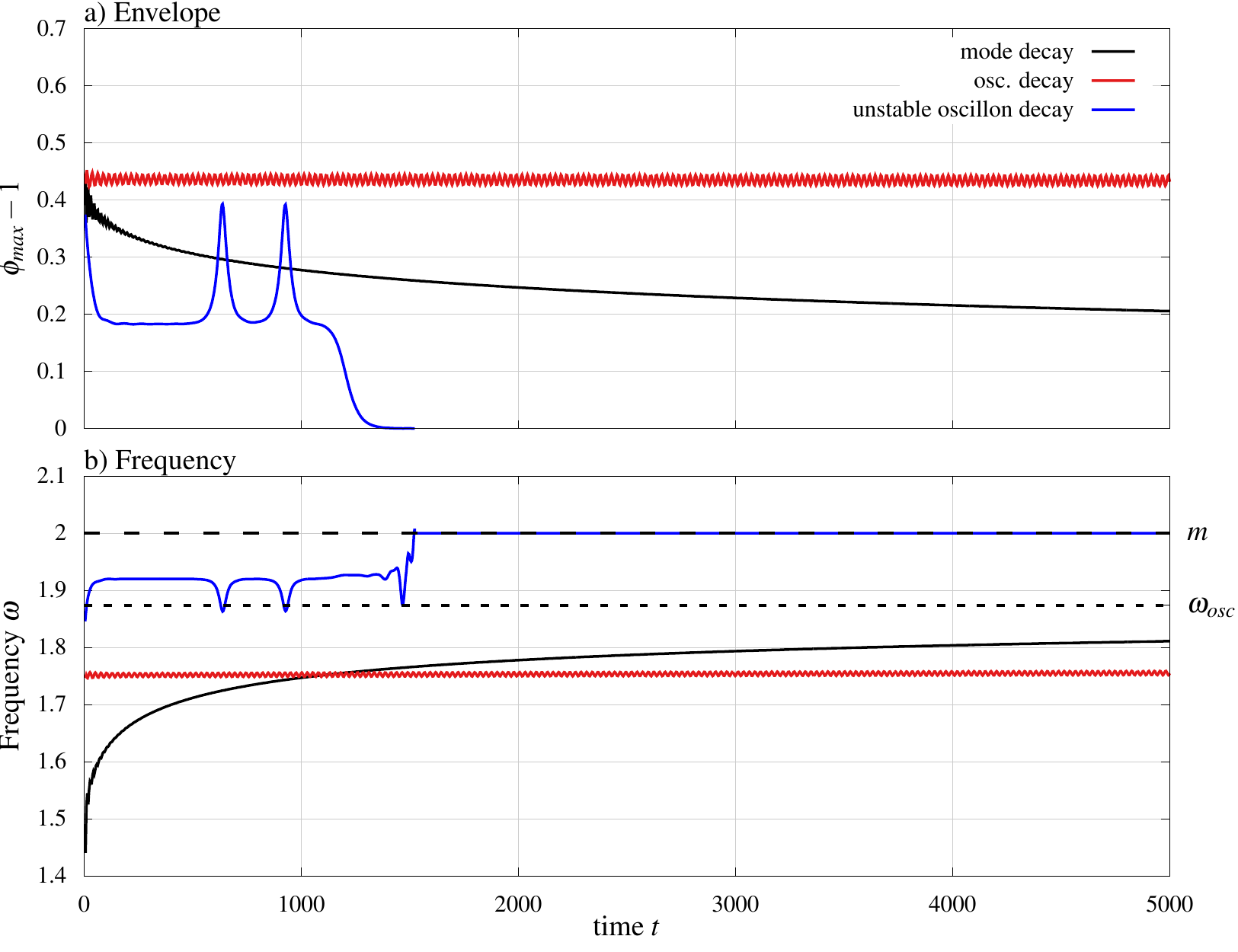}
\caption{\small  Possible relaxation scenarios for $A_0=0.6$. The plots show  the maximum envelope of oscillations at $x=0$
(upper plot) and the measured frequency as $\omega=2\pi/T$ (bottom plot), where $T$ is time between the subsequent maxima.
Decay of the oscillational mode (black curve); usual evolution of the oscillon configuration
(red curve);  decay of the unstable oscillon in the repulsive potential (blue curve).}\label{fig:relaxationLong1}
\end{center}
\end{figure}

The results of the numerical analysis are shown in the Fig.~\ref{fig:relaxationLong1}.
The initial data represent a symmetric profile $\phi(x,0)=1-A_0\sech^\alpha(x), ~\phi_t(x,0)=0$, where $\alpha $ is a positive
real parameter. Note that
in the presence of the potential in  Eq.~\re{linear-eq} with $V_0=\lambda(\lambda+1)$
the above initial data exactly match the profile of the corresponding oscillational mode when $\alpha=\lambda$.
In the absence of the potential, or in case of the repulsive potential, these initial data  evolve into the
oscillon solution.

The upper subplot of  Fig.~\ref{fig:relaxationLong1} shows the upper envelope of the
oscillations of the field $\phi(0,t)$.
The bottom subplot displays the basic frequency of the oscillations measured from the time period between
two subsequent maxima.
This observable is quite informative for identification almost periodic functions. However, in some situations
the basic frequency cannot be easily identified in this way because of numerical artifacts.
Thus we have also checked the results performing Fast Fourier Transform and generating the power spectrum of the data.

In  Fig.~\ref{fig:relaxationLong1} the black curve illustrates the evolution of the oscillational mode of the model
\re{linear-eq} for a particular choice of the potential parameter $V_0=1.19$ ($\lambda=0.7$).
Initially this mode is excited up to nonlinear regime setting $A_0=0.6$.
Due to the nonlinearity, the corresponding initial frequency ($\omega=1.5$) is much lower initially than its value
predicted from the linearized theory, $\omega_{osc}=1.8735$.
However, with time the energy is radiated away and the amplitude of oscillations slowly decays.
Then the frequency raises and it approaches the corresponding linearized value $\omega_{osc}$.
The decay rate of this mode follows the power law, $A\sim t^{-1/2}$ \cite{Manton:1996ex,Slusarczyk:1999iy}.

The red curve shows the evolution of the oscillon configuration,
which is produced  from the similar initial data setting $A_0=0.7$ in the absence of the external potential, $V_0=0$.
Note that both the amplitude and the frequency of the oscillations vary very slowly
in comparison to the oscillational mode. Some small modulations of the frequency are still visible.

The blue curve shows the evolution of the oscillon initial data with $A_0=0.7$ as above, however, now
in a presence of the repulsive potential, $V_0=-0.39296$.
As expected, for a given hight of the potential only the oscillons with amplitude larger then some critical value can exist.
Again, they slowly fade radiating energy away and we observe the decrease of the amplitude of the oscillations
due to radiation loss. At some point the nonlinearities cannot hold together the oscillon which starts to oscillate with highly modulated amplitude
(peaks of the blue line) just before it breaks in a short burst of radiation. This is the feature which has been observed for oscillons in higher
dimensions as well \cite{Gleiser:1993pt}.

For small positive values of $V_0$ it might happen that the oscillational mode would initially behave as oscillon (radiating very slowly) until the
linear
term $V_0(\phi-1)$ becomes large enough in comparison to the nonlinear term $(\phi-1)^2$.
The smaller $V_0$ is the smaller the coupling between the oscillational mode and the second harmonic. The linear modes (attractors of
the evolutions) become wider. The coefficient of proportionality multiplying the decay power law $t^{-1/2}$ becomes smaller.
There is a smooth transition between oscillational modes $V_0>0$ and the oscillons $V_0=0$.

For $V_0<0$ the oscillon does not have a supporting mode and the potential is repulsive. There exists a certain amplitude below which the oscillon
cannot exist as the nonlinearities have not enough strength to hold it together. In the evolution from the large amplitude at some point the oscillon
looses its stability as the amplitude reaches the critical value. The oscillon rapidly vanishes into the radiation. Sometimes the radiation can have
a form of two oscillons. In such a case the initial oscillon is literally torn apart by the potential into two smaller oscillons.

\subsection{Adiabatic transformation}
\begin{figure}[!h]
 \begin{center}
\includegraphics[width=.75\linewidth]{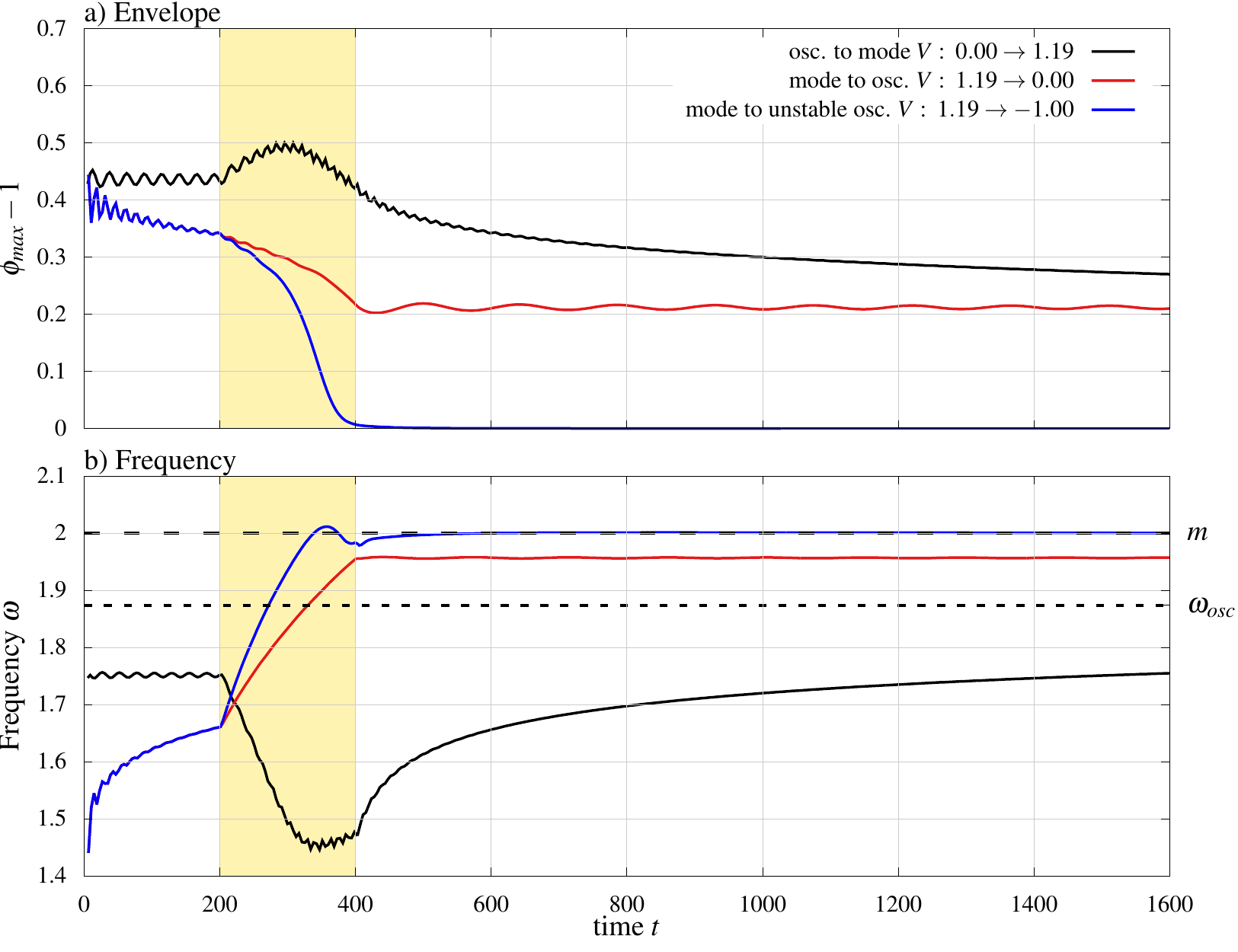}
\caption{\small  Examples of adiabatic transformations: (i)
the oscillon $A_0=0.6$ to the internal oscillational mode  (black curve);
(ii) the oscillational mode to the oscillon (red curve);
(iii) the oscillational mode to an unstable oscillon and its consequent decay (blue curve).}\label{fig:relaxationLong}
\end{center}
\end{figure}

An interesting evolution scenario was observed for the situation when the initial data set rapidly converges to the  oscillon
configuration in the absence of the external potential (black curve in  Figs.~\ref{fig:relaxationLong}).
Within the time interval $t \in [t_1, t_2]$ where $t_1=200$ and $t_2=400$
(yellow strip in Fig.~\ref{fig:relaxationLong}), we
gradually increase the depth of the potential as
\begin{equation}
 V(t) =V_0\frac{t-t_1}{t_2-t_1}.
\end{equation}
Here we set $V_0=1.19$ as above.
Then the amplitude of the oscillations increases and, as  $ t > t_2$, the oscillating state becomes trapped by the potential well.
Further, starting from this moment of time, the pattern of evolution of the configuration follows the usual
$t^{-1/2}$ law of the radiative decay of the oscillational mode. In other words, the oscillon state becomes smoothly transformed
into the oscillational mode.

The opposite transition is illustrated in  Fig.~\ref{fig:relaxationLong}
by the red curves. Initially, there is an
excited oscillational mode in the potential well with the same values of the parameters as above. As $t_1<t<t_2$ the potential is
turned off smoothly by setting
\begin{equation}
  V(t) = V_0\frac{t-t_2}{t_2-t_1},
\end{equation}
and the oscillation mode becomes transformed into the oscillon state with
amplitude  a bit above $A=0.2$ and the frequency $\omega=1.9566$. This is clearly below the mass
threshold. Both the frequency and the amplitude of the configuration are modulated by small oscillations,
however, the corresponding average values are almost constant.

Finally, the blue curves in Fig.~\ref{fig:relaxationLong} show the decay of the internal mode in the
potential with the initial depth $V_0=1.19$ as it changes adiabatically to $V_0=-1$ as $t_1<t<t_2$. During the change the oscillational mode turned
into an oscillon which was
destroyed by the repulsive potential as $V_0<0$.

\begin{figure}[!h]
 \begin{center}
 \includegraphics[width=.75\columnwidth]{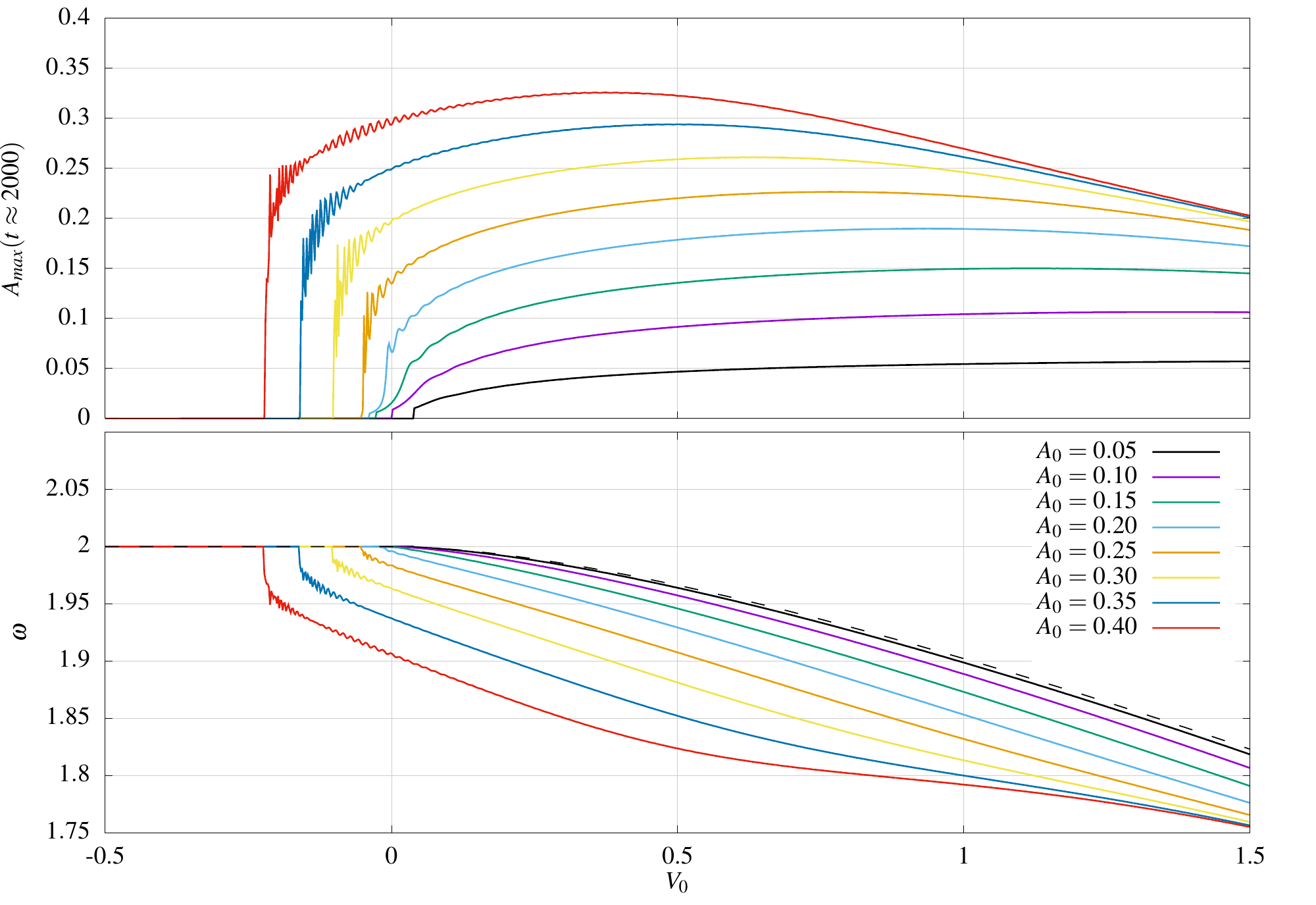}
\caption{\small Upper plot: Amplitude measured after about $t=2000$ from initial data with $\alpha=0.4$ as a function of $V_0$.
Bottom plot: Measured frequency of the final state. The dashed line represents
the frequency of the bounded mode of the corresponding potential at $A_0=0$.}\label{fig:FinalState}
\end{center}
\end{figure}

\subsection{Final state}
Let us now consider dependency of the relaxation scenarios of the initial data on the parameters of the input configuration
$\phi(x,0)=1-A_0\sech^\alpha(x),~ \phi_t(x,0)=0$, as above.
In Fig.~\ref{fig:FinalState} the upper plot shows the amplitude of the oscillations at $x=0$
measured after time interval $t=2000$ for different values of the initial
amplitude $A_0$ as a function of the potential depth $V_0$ with $\alpha=0.4$.

We observe, that for relatively small values of the initial amplitude, the oscillations above the vacuum are very small.
This suggests that the system evolves following the linearized dynamics.
The lower plot of Fig.~\ref{fig:FinalState} shows the measured frequency, which for small values of $A_0$ is very
close to the frequency of the bounded mode of the corresponding linearized system. Thus, the initial perturbation
can be decomposed into the linearized modes: one bounded mode and the modes from the continuum.
The scattering modes move
away from the center where only the oscillational mode remains.
Note that as the initial amplitude increases, the frequency of the oscillations decreases.

Interestingly, there is a region $0>V_0>V_{cr}$, at which the final state is still oscillating, even in the case of the
repulsive potential. Clearly, the corresponding frequencies are still below the mass threshold,
although the repulsive potential does not support an oscillational mode.

In  Fig.~\ref{fig:FinalState} one can see some modulations of the amplitude but, in
general, there is no evidence of a clear
distinction at the boundary $V_0=0$, both on the upper and on the lower plots.

It is usually assumed that whenever such persistent oscillations with the frequency below
the mass threshold are seen in the absence of oscillational
mode, an oscillon is produced from the initial data \cite{Copeland:1995fq}.
From the above simulations we can see that a distinction between an oscillon and an oscillational mode
is quite artificial. The only difference is that the frequency of the oscillational modes
tend to the frequency, calculated from
the linearized model, as the amplitude decreases. In the case of oscillons, there is no mode, toward which
they could approach, so they just slowly decay into the vacuum $V_0=0$,
or until the repulsive potential with $V_0<0$ destroys the oscillon.

\subsection{Phase diagram}
\begin{figure}[!h]
 \begin{center}
 \includegraphics[width=.75\linewidth]{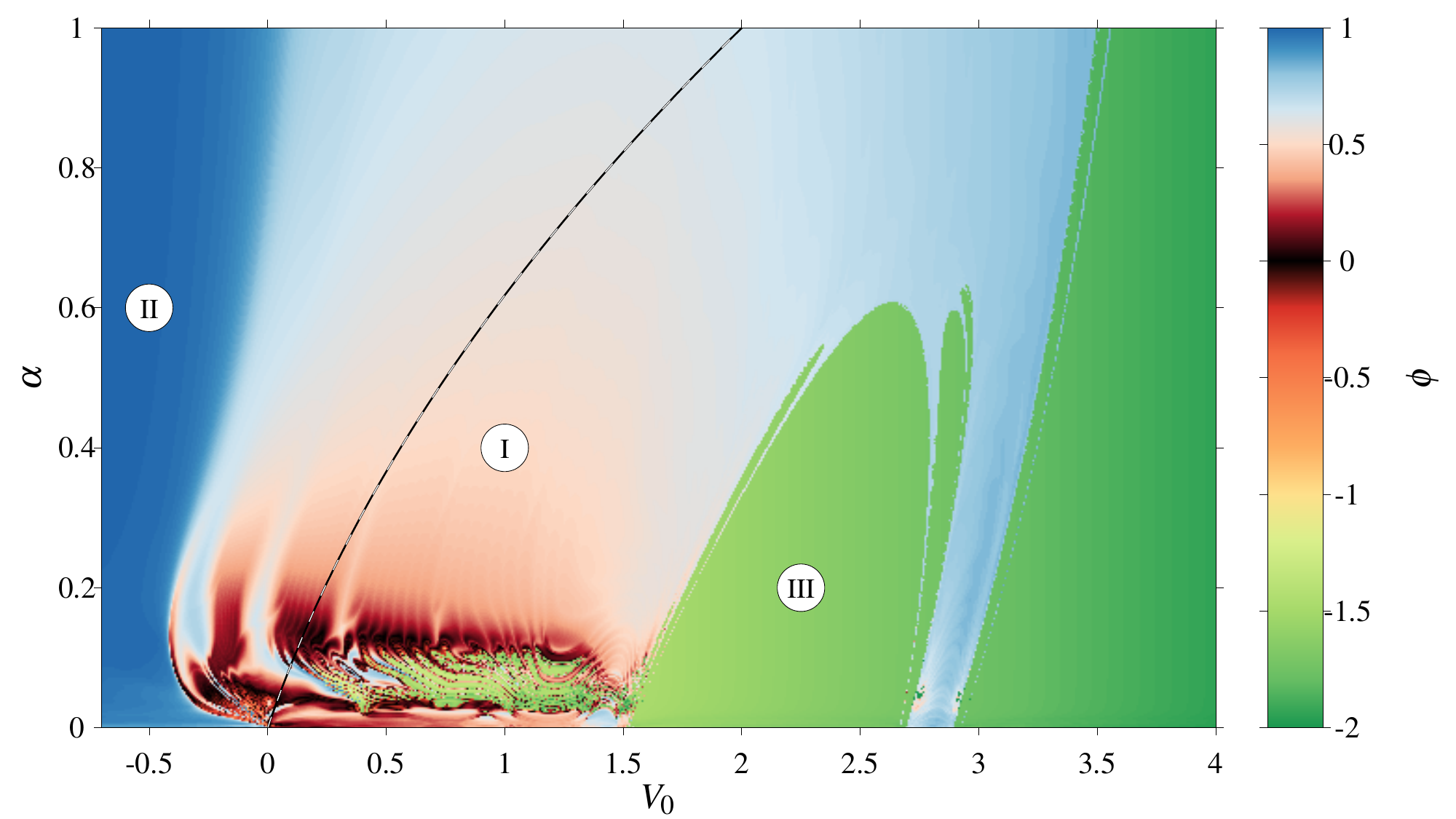}

\caption{\small Minimal value of $\phi(0,t)$ for $80<t<100$ for $A_0=0.4$.
The dashed curve corresponds to the parameters of the exact solution for the internal mode in the potential $V(x)$.
White-red regions (I) correspond to oscillational mode (if $V_0>0$) and
well developed stable oscillon (if $V_0\leq0$), dark blue (II) correspond to unstable oscillon, yellow-green (III) kink antikink
creation.}\label{fig:scan1}
\end{center}
\end{figure}

\begin{figure}[!h]
 \begin{center}
 \includegraphics[width=.75\linewidth]{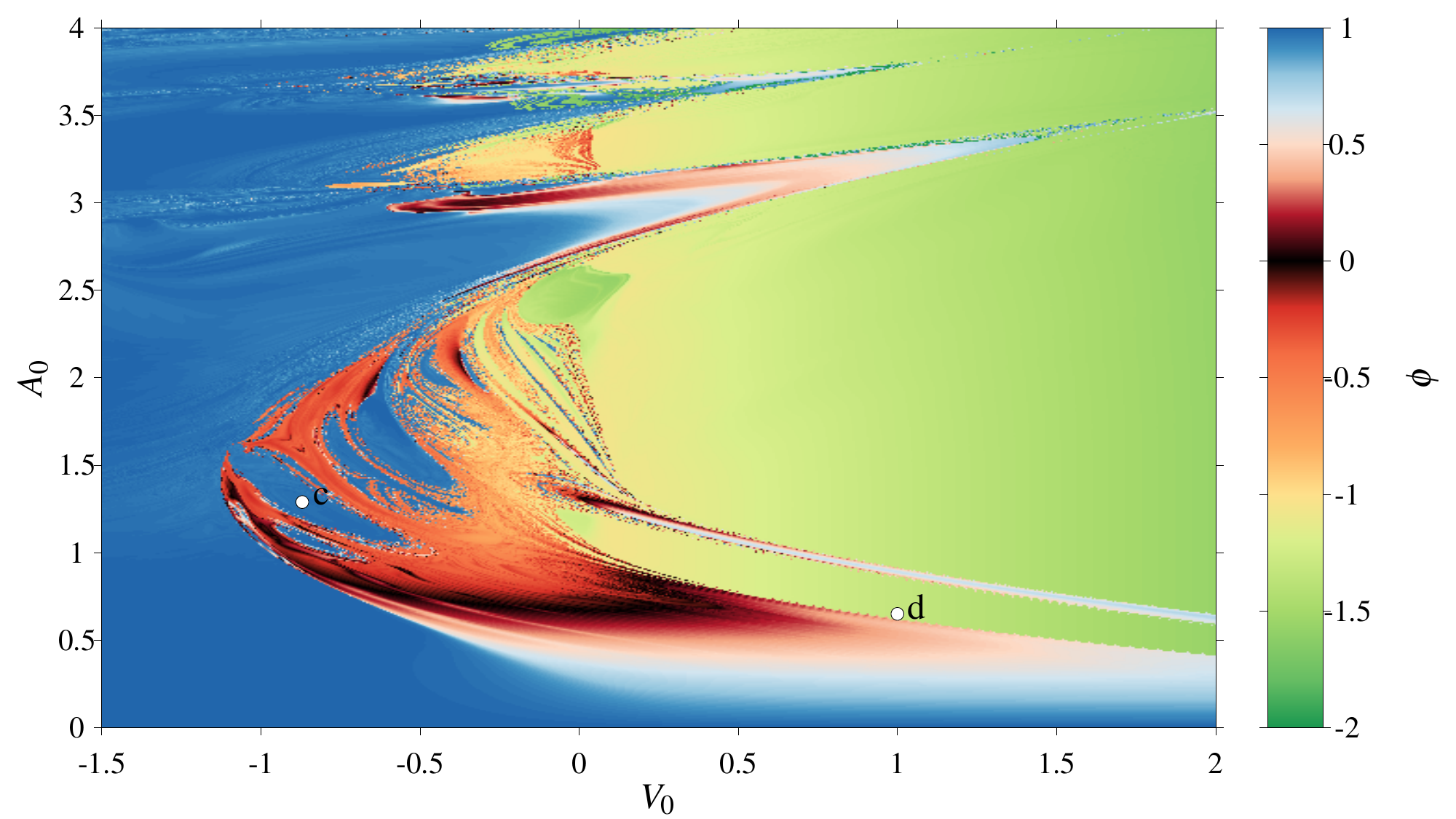}
\caption{\small Minimal value of $\phi(0,t)$ for $\alpha=0.4$.}\label{fig:scan2}
\end{center}
\end{figure}

In Fig.~\ref{fig:scan1} we display a \textit{phase diagram}
indicating regions with different scenarios (basins of attraction) described above.
The color scheme represents the minimal value of the field $\phi(0,t)$ at the center of the trapping potential
for the time interval $80<t<100$ as function of the parameters $\alpha$ and $V_0$.
The dashed curve corresponds to the internal mode of the corresponding potential $V(x)$, $\alpha=\lambda$.
Similar diagram, but now depending on the parameters $A_0$ and $V_0$ for fixed value $\alpha=0.4$ is shown
in Fig.~\ref{fig:scan2}.

\begin{figure}[!h]
 \begin{center}
 \includegraphics[width=.75\linewidth]{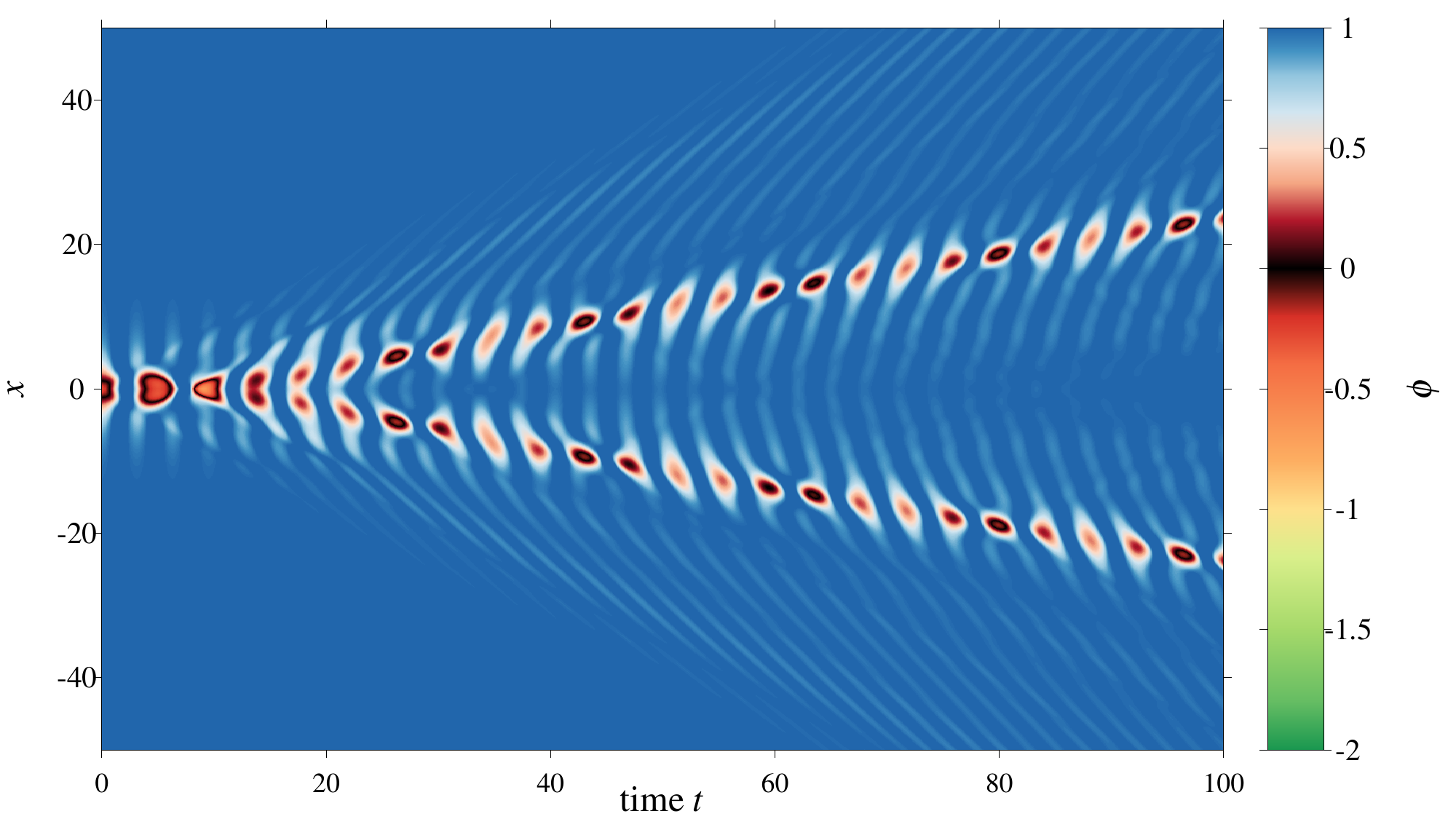}
\caption{\small Unstable oscillon torn by the external potential into two oscillons ($V_0=-0.81$, $A_0=1.32$).}\label{fig:scan3}
\end{center}
\end{figure}

Firstly, we observe that for negative values of the parameter $V_0$, which correspond to region (II) colored in blue,
the minimal value of the field
approaches the vacuum. Physically, the corresponding initial oscillon configuration, which rapidly decays into
the burst of radiation or a pair of
smaller amplitude oscillons, see Fig.~\ref{fig:scan3}.

\begin{figure}[!h]
 \begin{center}
 \includegraphics[width=.75\linewidth]{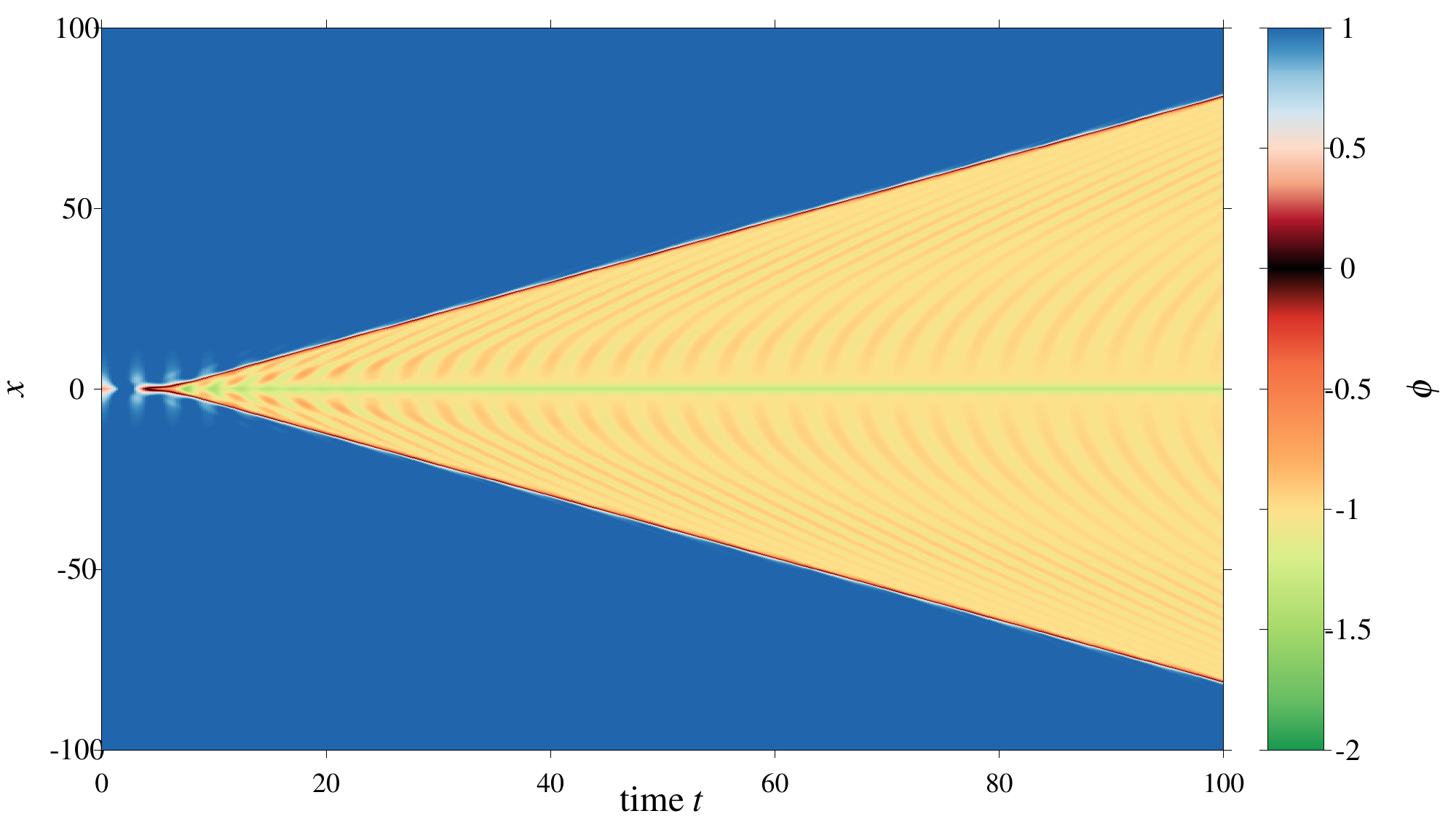}
\caption{\small $K\bar K$ production ($V_0= 1.06$, $A_0=0.69$).}\label{fig:scan4}
\end{center}
\end{figure}

For sufficiently large value of the parameter $\alpha$,  which defines the width of the input data,
and relatively small positive values of the parameter $V_0$, we observed that the initial configuration
evolves into slowly radiating oscillating mode, this scenario corresponds to the pink-white regions (I) in this plot.
Note that as $\alpha$ becomes relatively small, the amplitude of the oscillations of the final state is increasing and the dynamics
of the field turns out to be chaotic with multiple resonances related with production and annihilation of the kink--anti-kink pairs (region (III)
see Fig.~\ref{fig:scan4}).
On the other hand, this domain smoothly extends to the region, where $V_0<0$ and the internal mode ceases to exist. Thus, in such a
case the final state represents an oscillon.

\section{Conclusions}

The main objective of this work is to demonstrate that there is an intrinsic relation
between the oscillational normal modes and the oscillon configurations, which, under
certain assumptions can be smoothly transformed into each other.
As a simple  example model we considered a modification of the 1+1 dimensional $\phi^4$ theory with a symmetric potential, which was used to
introduce an external perturbation of the spectrum.

In our numerical experiment the potential was changed adiabatically from its attractive from, which supports an
oscillational mode, to repulsive interaction. In the course of the evolution we observe the transformation of
an oscillational mode into an oscillon configuration. In the opposite case,  adiabatically switching the potential
from repulsive to attractive form, we observe that an oscillon becomes smoothly transformed into
an excited oscillational mode, which later starts to decay obeying the corresponding power law.

We also studied various evolution scenarios of a certain class of initial data.
We found that, depending on the depth of the potential and the profile of the initial data, the resulting configuration
may represent either the oscillons or the oscillational mode.
We also conjecture that the distinction between the oscillational mode and oscillon is rather artificial and in
moderately long time dynamics there
is no clear distinction between the two.
In our phase diagrams we were unable to see any distinctive boundary separating oscillon from oscillational
mode evolution types.
Only extremely long time evolution could show the difference in the limiting case $V_0\to 0$.
Then the oscillational modes would vanish oscillating with the final frequency below the mass threshold and
amplitude tending to zero while the
oscillons either with frequency tending to the mass threshold and vanishing amplitude ($V_0=0$ case) or
with some nonzero critical amplitude
(repulsive $V_0<0$ case).

We did numerical simulation considering different values of the parameters of the potential $b$ and $V_0$, which
corresponded to higher number of (even) bounded modes.
In the most cases we observed that the lowest mode was an attractor in the time evolution of the initial data.


In the Appendix we studied the radiative decay of a single oscillational mode.
We checked the validity of the Manton-Merabet approach, which leads to the power law decay of the
mode, by direct measuring of the amplitude of the radiation, emitted by the excited mode.
This approximation, however, fails in the case of deeper oscillational modes, in such  a case higher harmonics
are hidden below the mass threshold and the mode decays much slower than it is predicted by the Manton-Merabet law
\cite{Dorey:2015sha}.
One of the most interesting findings we examined in this paper is the observation of burst
or radiation, which happens as the slowly decreasing
frequency of higher harmonic of the initially highly excited oscillational mode, crosses
the mass threshold and starts to propagate.
This feature may be important in other nonlinear field theories, which may support deeper
oscillational modes, for example in the Skyrme model \cite{Adam:2017czk}.

Another important observation is that as the potential vanishes, the frequency of the oscillational mode
tends to the frequency of the mass threshold from
below. Moreover, the radiation amplitude calculated from the second order perturbative scheme also vanishes
along with the potential.
This yields yet another illustration of smooth character of the transition from the oscillational mode to the
oscillon in the parametric space.

Our observations could indicate that similar scenario is also possible in higher dimensional field theoretical systems.
In principle one does not need to limit the considerations to the external potential.
Topological defects or solitons can be the source of a very similar potential as examined in the current paper.
Both oscillons and the bound modes of the solitons were studied extensively in the past.
However, an open question remains whether a configuration of a topological defect bound with an oscillon can exist.

Another question, which we hope to be addressing in the near future, is to consider the dynamics of the kinks in the
presence of the localizing potential. It will extend the related analysis of the dynamics of the solitons with delta-like inhomogeneity
\cite{Fei:1992dk,Goodman,Kivshar:1991zz}.

\noindent{\bf Acknowledgements:} The authors are grateful to Patrick Dorey for inspiring and valuable discussions and
Boris Malomed for his valuable comments and correspondence.
YS  gratefully acknowledges support from the Russian Foundation for Basic Research
(Grant No. 16-52-12012), the Ministry of Education and Science
of Russian Federation, project No 3.1386.2017, and DFG (Grant LE 838/12-2).
YS would like to thank the Institute of Physics, Jagiellonian University, Krakow
for its kind hospitality.

\appendix
\section{Time evolution of the oscillational mode}
\subsection{Manton-Merabet power law}
Let us assume that a mode around the trivial vacuum $\phi=1$ is excited.
The linear analysis gives us the frequency of its small amplitude oscillations, it corresponds to the
solution of the Eq.~(\ref{eq:linsol}).
In nonlinear theories, large amplitude excitations can change the fundamental frequency of the oscillations.
Oscillational modes are unstable, they radiate due
to the coupling to the modes of continuum, so the frequency of oscillations  and the amplitude of the
mode $A$ evolve with time, the amplitude of the oscillations will slowly decrease while the corresponding
frequency (usually) will increase towards the value obtained from the linearization.

In a similar consideration of the time evolution of the internal oscillational mode of the
usual $\phi^4$ kink, Manton and Merabet showed that this  mode decays due
to the radiation following the rule $A\sim t^{-1/2}$  \cite{Manton:1996ex}.
Their arguments can be summarized as follows.
The evolution of the mode can be described via a perturbative expansion in the powers of the amplitude
\begin{equation}
 \phi=1+\sum_{n=1} A^n\xi^{(n)}.
\end{equation}
Substituting this expansion into the linearized field equation for fluctuations, we can see that
each term of the perturbation series should satisfy the equation of the form:
\begin{equation}
 \xi^{(n)}_{tt}+\mathbf{L}\xi=g^{(n)}\left(\xi^{(1)}, \ldots, \xi^{(n-1)}\right) \, ,
\end{equation}
where
\begin{equation}
 \mathbf{L}=-\partial^2_{x}+m^2+V(x) \, ,
\end{equation}
is a linear operator and
\begin{equation}
 g^{(n)}\left(\xi^{(1)}, \ldots, \xi^{(n-1)}\right) \, ,
\end{equation}
is a source term which depends only on the solutions of the lower order.

For example in the first two orders we obtain
\begin{equation}
 g^{(1)}=0,\qquad g^{(2)}=-6{\xi^{(1)}}^2
\end{equation}
Note that the in the second order the source term is given by the square of the solution of the first order
equation. If the time dependence of the fluctuations in the first order is harmonic, i.e.
\mbox{$\xi^{(1)}\sim\cos\omega t$}, the source term oscillates like \mbox{$g^{(2)}\sim 1+\cos(2\omega t)$}.
This
is a source of radiation with a frequency $2\omega $,
which carries away the energy from the mode causing its decay.

Note that the amplitude of the radiation in
the lowest order is proportional to $A^2$, thus the energy flow is $dE/dt\sim A^4$.
Further, since the energy of the oscillational mode is proportional to $A^2$ the
decay of the mode should follow the power low
\begin{equation}
\frac{dA}{dt}\sim -A^3\Rightarrow A(t)\sim t^{-1/2} \, .
\end{equation}
Here the proportionality coefficient can be found using Green
function technique and expressed as an integral over a function of eigenfunctions
of the corresponding linearized problem:
\begin{equation}
 \xi^{(2)}(x\to\infty)\approx \frac{3e^{2i\omega
t}\eta_{-k}(x)}{2W}\int_{-\infty}^{\infty}\eta_0(x')^2\eta_k(x')\,dx'+c.c\equiv\mathcal{A}\cos(2\omega t
-kx+\delta).
\label{eq:Green}
 \end{equation}
where $W=\eta_k\eta'_{-k}-\eta'_k\eta_{-k}$ is a Wronskian and $k=\sqrt{4\omega ^2-m^2}$ is the wave number for the second harmonic,
$\mathcal{A}$ is the amplitude of the radiation and $\delta$ is some phase shift.

In general, the eigenfunctions  of the P\"oschl-Teller potential can be expressed  in terms of Legendre polynomials
\begin{equation}
 \eta_k(x)=P^{ik}_\lambda(\tanh x), \qquad k=\sqrt{k^2-m^2} \, ,
\end{equation}
however, the corresponding  integrals in \re{eq:Green} can be calculated analytically
only for certain values of $\lambda$. As $\lambda\to0$ (or $V_0\to0$) the amplitude of the second harmonic also vanishes. This is yet another
similarity with the oscillons. In \cite{BOYD1995311} it was shown that oscillons radiate beyond all orders, meaning that the radiation amplitude
cannot be represented as a simple power series and that all coefficients of the perturbation series are equal zero.

\subsection{Numerical verification}
\begin{figure}[!ht]
 \begin{center}
\includegraphics[width=.75\linewidth]{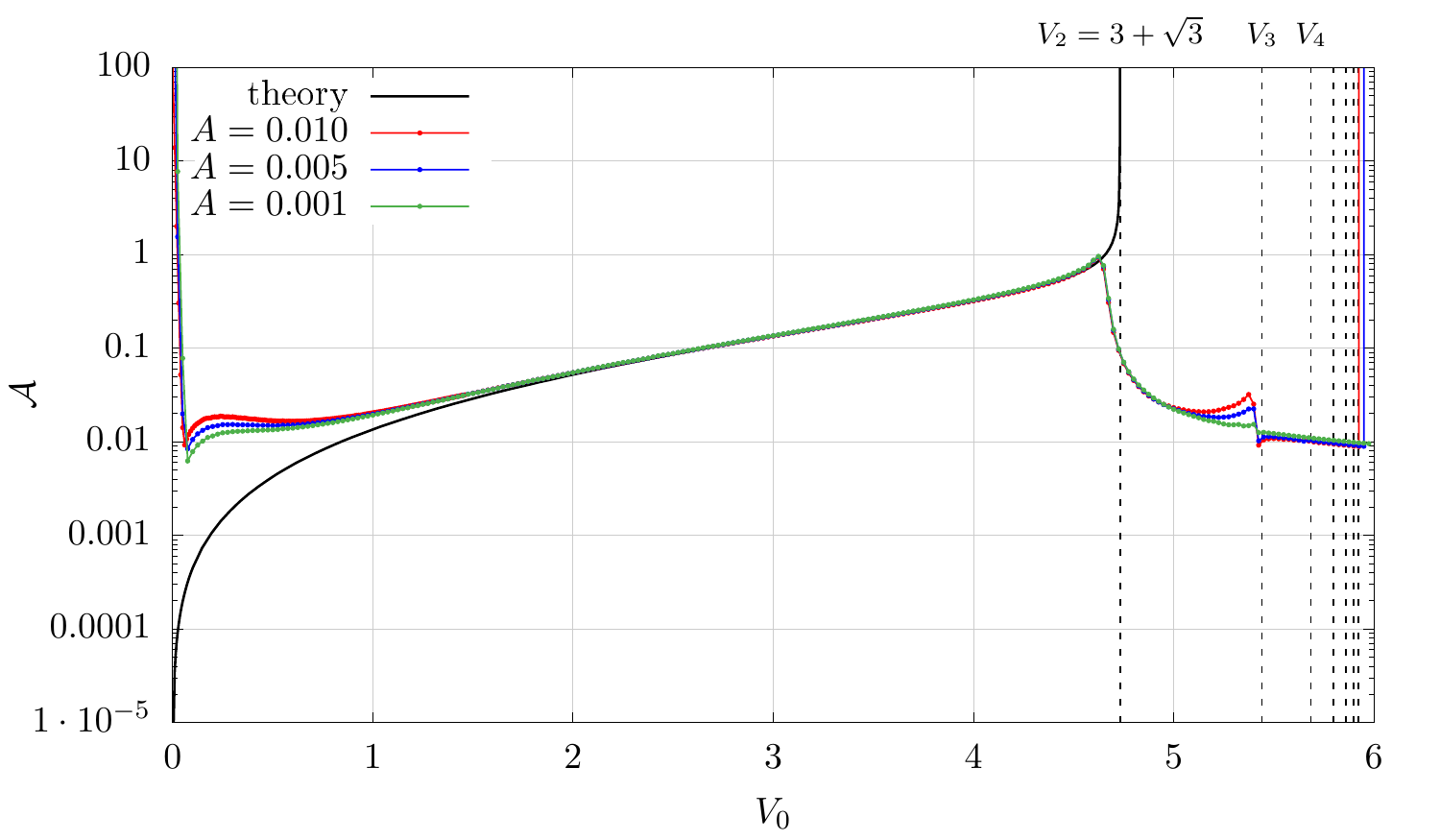}
\caption{\small Radiation measured at $x=200$ at $t=1000$ (divided by $A^2$) compared with the theoretically predicted values (\ref{eq:Green})
(black curve).
}\label{fig:Radiation}
\end{center}
\end{figure}

We have calculated the amplitude numerically and compared it with the measurements performed for small
values of amplitude $A$ of the oscillational mode (Figure~\ref{fig:Radiation}). The amplitude was measured at the distance $x=200$ at the time
$t=1000$.
The amplitude obtained from evolution is in good agreement with the second order calculations for $1.5<V_0<4.6$.
Below the lower value the amplitude of the
second harmonic is small and comparable with other types of radiation.
Below $V_0=1$ the discrepancy is significant but the scaling $\mathcal{A}\sim A^2$ seems to be preserved.
This could mean that either our predicted amplitude is wrong or other modes dominate.
In our simulations we started the evolution with initial conditions corresponding to the excitation of the pure oscillational mode at a
given time $t=0$.
The source term $g^{(2)}$ is then not entirely harmonic (with a single frequency). It should be multiplied by a Heaviside step function
\footnote{more precise analysis of this problem  in the case of excited $\phi^4$ kink can be found in \cite{Romanczukiewicz:2003tn}}.
This means that all the frequencies ($\mathcal{F}[H(t)e^{i\omega_0t}]=\frac{1}{2}\delta(\omega-\omega_0)-\frac{i}{2\pi(\omega-\omega_0)}$)
contribute to the evolution and a single frequency approximation fails.
As the result, the radiation amplitude consists of not only the amplitude of the second harmonic \re{eq:Green} but is a sum/integral of amplitudes of
all the modes. This should be especially visible when the second harmonic amplitude tends to 0 as $V_0\to 0$.

\begin{figure}[!ht]
 \begin{center}
\includegraphics[width=.75\linewidth]{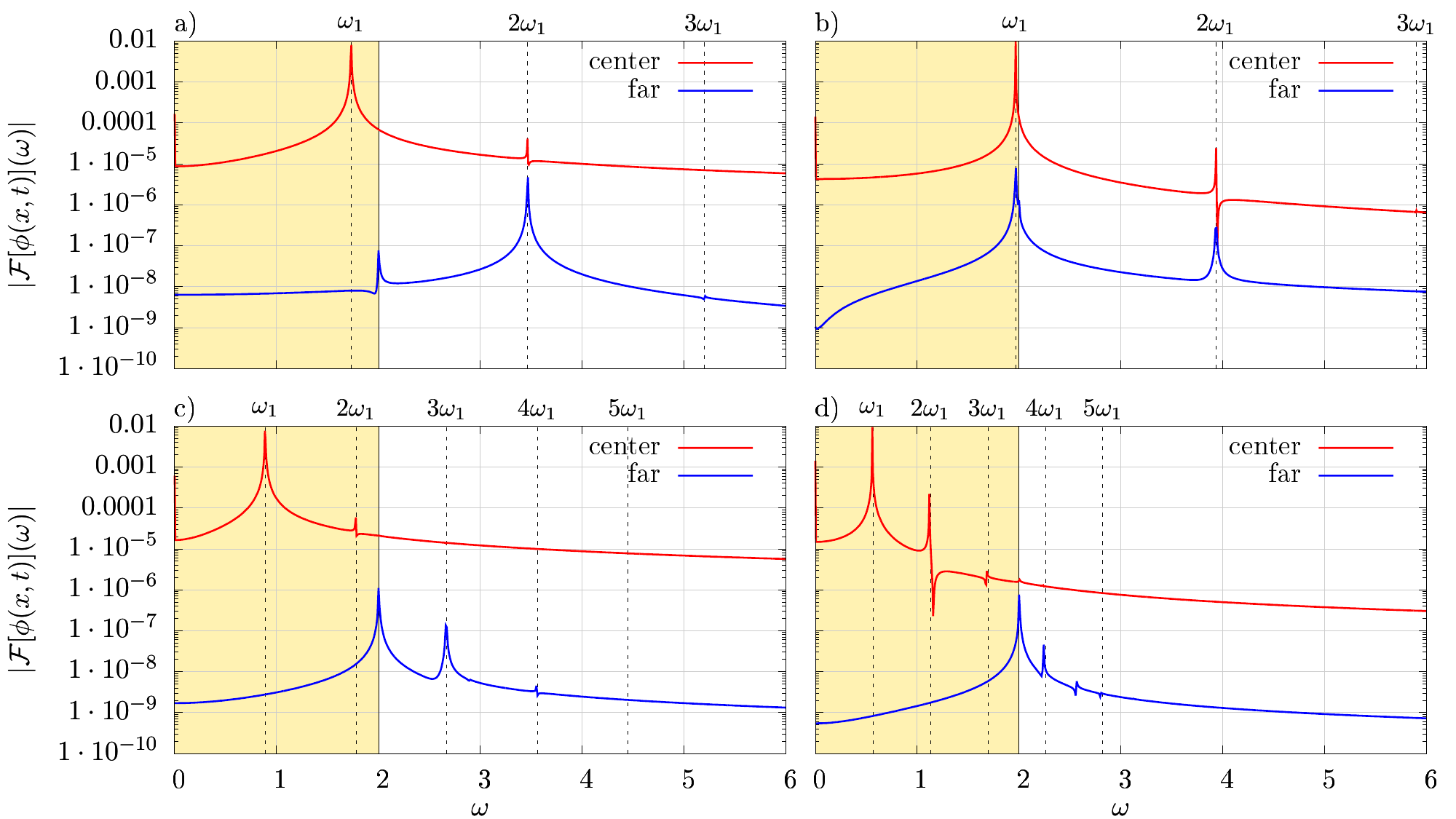}
\caption{\small Power spectra of the measured at the center $x=0$ (red) and far $x=200$ field  (blue) for different values of the potential 
depth $V_0$
and for initial excitation of the oscillational mode $A=0.01$. Four different scenarios correspond to a) $V_0=2.0$ - almost pure second harmonic, b) 
$V_0=0.5$ - large contamination of the near-the-threshold modes, c) $V_0=5.0$ - decay through the third harmonic, d) $V_0=5.6$ - decay through the 
fourth harmonic.}\label{fig:RadiationFFT}
\end{center}
\end{figure}

In order to verify the above predictions we have plotted a power spectra of the field measured at the center ($x=0$) and far away ($x=200$) from 
the
center for initially excited $A=0.01$ oscillational mode for a few values of $V_0$ (Figure \ref{fig:RadiationFFT}).
For $V_0=2$ (a) the most dominating frequency in the far field measurements is $2\omega_1$. The height of a peak is not much lower than the
predicted $\mathcal{A}A^2=5.2\cdot10^{-6}$. For $V_0=0.5$ (b) the height for the second harmonic is also close to the predicted value of
$4.2\cdot10^{-7}$ but it is not the most dominating frequency now. The spectrum is highly contaminated with the frequencies near the mass threshold
$\omega\approx m=2$. So that means that indeed the radiation amplitude is proportional to the square of the amplitude of the oscillational mode,
but for $V_0\to 0$ the second harmonic is not the most dominating frequency in the spectrum.

\subsection{Deep modes relaxation}
Note that the  above consideration is valid only when two following conjectures are fulfilled.
First, we consider the amplitude of oscillations to be small enough to secure the convergency of the
perturbation series in $A$. Another assumption is that the second harmonic belongs to the continuous
scattering spectrum. This second conjecture, however, is true only when the potential is not too
deep so that
\begin{equation}
 b^2\lambda^2<\frac{3m^2}{4}\, ,
\end{equation}
or, in other terms:
\begin{equation}
 V_0<\frac{3m^2}{4}+\frac{m\sqrt{3}}{2}b \, .
\label{eq:secondbound}
\end{equation}
If this condition is not fulfilled, the second harmonic remains below the mass threshold and
cannot propagate. This happens for $V_0>V_2=\sqrt{3}+3$ (for $b=1,\;m=2$).
For $V_0=V_2$ the integral in (\ref{eq:Green}) is divergent  (Figure~\ref{fig:Radiation}). For $V_0>V_2$ the approximation (\ref{eq:Green}) is not
valid.
Numerical simulations show that even for some range below $V_2$ the perturbation approach fails. This corresponds to the case when the 
frequency of the second harmonic is very close to the mass threshold ($2\omega_1$ is almost in a resonance with $m$).

In order to calculate the energy loss of the oscillational mode for $V_0>V_2$, one has to include the higher order terms of 
perturbation series in consideration.
Assuming that the third harmonic is the first one, which can propagate, we can expect that the
decay rate of the mode becomes
\begin{equation}
 A(t)\sim t^{-1/4}\, .
\end{equation}
However, as the potential depth is increasing, more and more higher harmonics move down
below the mass threshold. This happens as
\begin{equation}
 \lambda_n=\frac{m}{nb}\sqrt{n^2-1}\,,\qquad V_n=\lambda_nb^2(\lambda_n^2-1)\,,
\end{equation}
where $n\ge2$ is the number of the first propagating harmonic. It yields the generalized power
decay law of the oscillational mode
\begin{equation}\label{eq:powerdecay}
  A(t)\sim t^{-\frac{1}{2n-2}}\,.
\end{equation}
This feature was first pointed out in \cite{Dorey:2015sha}. Power spectra exemplifying the radiation through the third and forth harmonics are shown
in the Figure \ref{fig:RadiationFFT} c) and d) respectively.

Note that in the consideration above we assume that the amplitude of the mode  $A$ is small enough, so
the higher order corrections do not contribute much.

The nonlinearities can act at least in two ways. One is the appearance of that higher harmonics as considered
earlier.
The other effect is that the measured frequency $\omega(A)$ is lower than the eigenfrequency $\omega_1$ obtained from the linear analysis, usually:
$\omega(A)=\omega_1-c_2A^2+\cdots$, with $c_2>0$.
In the classical mechanics this effect can be seen in the case of pendulum or anharmonic oscillator.
Let us consider a mode with $\omega_1$ just above $m/2$. When the mode is largely excited, it may happen that $\omega(A)<m/2$.
In this scenario, the radiative decay would occur in two stages.
Initially the second harmonic with the frequency $2\omega(A)<m$ cannot propagate. The mode decays slowly through the third harmonic.
As the amplitude decreases, the frequency $\omega(A)$ grows and, for certain amplitude, crosses $m/2$, allowing the second harmonic to
propagate. As the result the relaxation process accelerates. At some distance this transition can be seen as a sudden burst of radiation.

\begin{figure}[!h]
 \begin{center}
\includegraphics[width=.75\linewidth]{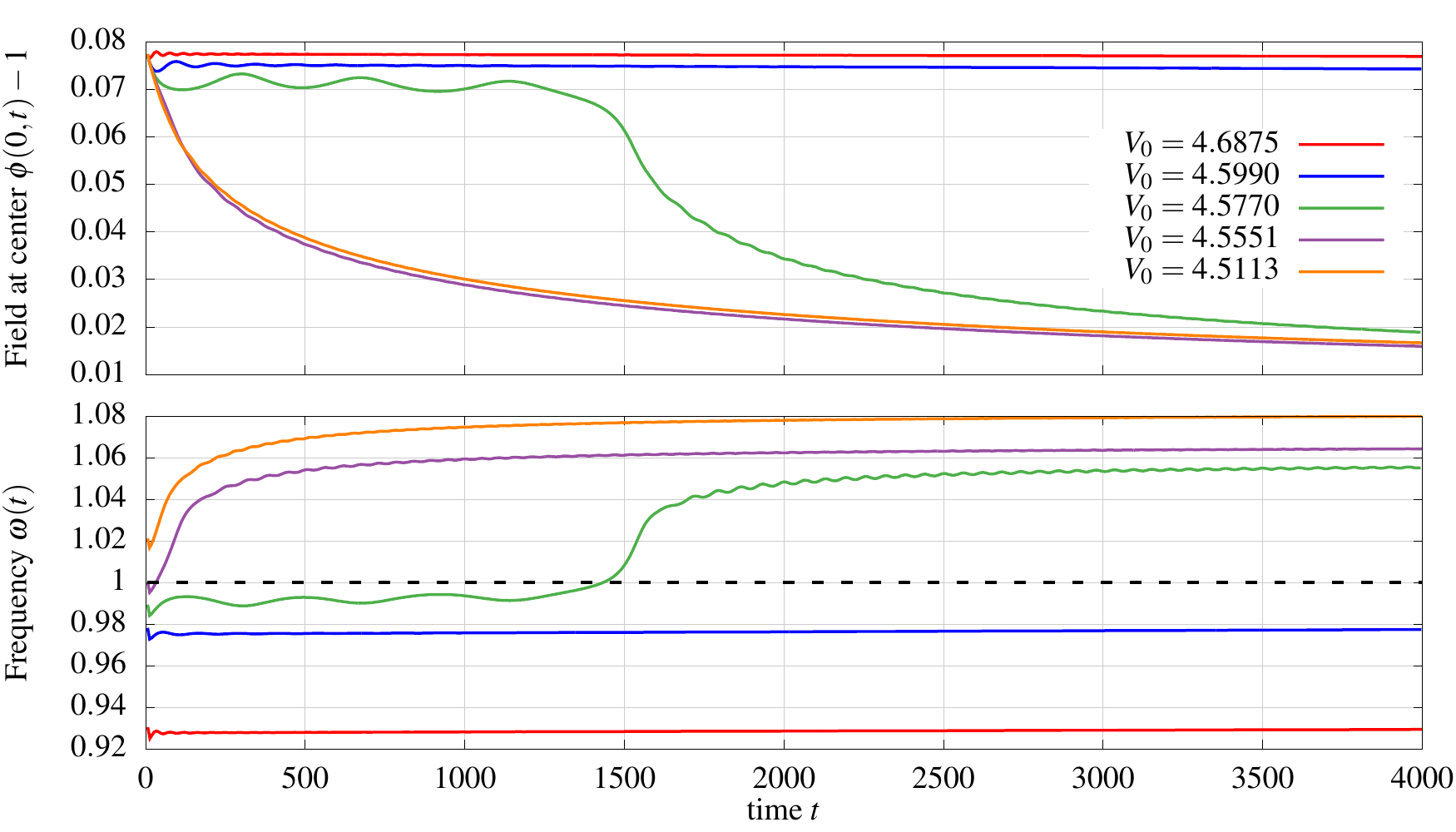}
\caption{\small Time evolution of the oscillational mode with amplitude $A_0=0.08$ potential width $b=1$ and various depths
just below the value when the second harmonics is hidden in the linear approximation.
Upper plot shows the maxima of field value at $x=0$, the bottom plot shows
the measured frequency $\omega(t)=2\pi/T$, where $T$ is a time between two subsequent maxima.
The red and
blue curve represent slow relaxation through the third harmonic,
the green curve shows the transition between two different types of decay,
which appear due to the nonlinearities related with relatively large values of the amplitude,
the purple and orange curves correspond to the radiative decay through the propagation of the
second harmonics.}\label{fig:relaxationMode}
\end{center}
\end{figure}

In Fig.~\ref{fig:relaxationMode} we show different scenarios of decays of the oscillational mode. In our numerical simulations
we set $b=1$ and we suppose that the potential depth is restricted by the inequality
\re{eq:secondbound}, i.e.
\mbox{$V_0\lesssim V_2=3+\sqrt{3}\approx4.7321$}. Thus, in the (close to) linear regime, $A\to 0$, the first harmonic is initially bounded, while the
second harmonic can propagate.

The initial data were taken according to \re{eq:linsol} as
\begin{equation}
\phi(x,0)=1+\frac{A_0}{\cosh^{\lambda}(bx)}\, , \qquad\phi_t(x,0)=0.
\end{equation}
The value of the initial amplitude $A_0=0.08$ is just enough to lower the frequencies of the oscillational modes below the second frequency threshold
 $\omega(A_0, V_0)<m/2=1$ for $V_0\geq4.5770$. The relaxation of these modes is much slower than the relaxation of the modes, for which the frequency
is above  the threshold $\omega(A_0, V_0)>m/2$ for $V_0<4.5770$.

In Fig.~\ref{fig:relaxationMode} we also show an example of the transition between two different types of the decay,
when initially the frequency was set below $m/2$
but in time the sufficient amount of energy was radiated so
that the frequency would increase above the threshold $m/2$ releasing the second harmonics. As the result the energy loss of the
oscillational mode due to its radiative decay increased dramatically.

Note that for the field of the decaying oscillating state (the purple and orange curves in the
Fig.~\ref{fig:relaxationMode}) for $t>1000$  can be nicely fitted by the function
\begin{equation}
 \phi(0,t)=1+a(t-t_0)^c
\end{equation}
where in both cases $c=-0.478$. Thus, the prediction of the power law decay is correct.

\bibliography{perturbedOscillon}
\bibliographystyle{JHEP}

\end{document}